\documentclass[10pt,twocolumn,aps,prl,superscriptaddress,amsmath]{revtex4-1}

\usepackage{graphicx}
\usepackage{hyperref}
\usepackage{bbold}

\begin{document}


\title{Origin of bulk uniaxial anisotropy in zinc-blende dilute magnetic semiconductors}


\author{M.\ Birowska}
\affiliation{Institute of Theoretical Physics, Faculty of  Physics, University of Warsaw, ul.\ Ho{\.{z}}a 69, PL-00-681 Warszawa, Poland}

\author{C.\ {\'{S}}liwa}
\affiliation{Institute of Physics, Polish Academy of Sciences, al.~Lotnik\'ow 32/46, PL-02-668 Warszawa, Poland}

\author{J.\ A.\ Majewski}
\affiliation{Institute of Theoretical Physics, Faculty of  Physics, University of Warsaw, ul.\ Ho{\.{z}}a 69, PL-00-681 Warszawa, Poland}

\author{T.\ Dietl}
\affiliation{Institute of Theoretical Physics, Faculty of  Physics, University of Warsaw, ul.\ Ho{\.{z}}a 69, PL-00-681 Warszawa, Poland}
\affiliation{Institute of Physics, Polish Academy of Sciences, al.~Lotnik\'ow 32/46, PL-02-668 Warszawa, Poland}


\date{\today}

\begin{abstract}
It is demonstrated that the nearest neighbor Mn pair on the GaAs (001) surface has a lower energy for the $[\bar{1}10]$ direction comparing to the [110] case. According to the group theory and the Luttinger's method of invariants, this specific Mn distribution results in bulk uniaxial in-plane and out-of-plane anisotropies. The sign and magnitude of the corresponding anisotropy energies determined by a perturbation method and {\em ab initio} computations are consistent with experimental results.
\end{abstract}


\maketitle


One of the founding blocks of condensed matter physics is the virtual crystal approximation \cite{Nordheim:1931_AP1,Nordheim:1931_AP2} allowing to extend the outcome of the group theory for a given crystal symmetry to alloys with a random distribution of their constituents. However, it has been recently demonstrated, combining the progress in {\em ab initio} simulations and nanocharacterization methods, that the condition of the random distribution is violated in a number of alloys, leading to striking consequences. In particular, it has been found that open $d$ shells of transition-metal (TM) cations diluted in non-magnetic compounds not only provide localized spins but also, through hybridization with band states, contribute significantly to the cohesive energy, particularly if TM atoms occupy neighboring sites \cite{Sato:2010_RMP,Bonanni:2010_CSR}. The resulting attractive force between magnetic cations leads to their aggregation, either triggered by appropriate post-growth high-temperature annealing, as found for (Ga,Mn)As \cite{De_Boeck:1996_APL}, or occurring at the growth surface during the epitaxial process, the case of (Ga,Fe)N \cite{Navarro:2011_PRB}.  The TM aggregation invalidates the main premise of dilute magnetic semiconductor (DMS) physics, namely that concerning the random distribution of TM spins over cation sites. One striking consequences is the appearance of surprising high temperature ferromagnetism in numerous magnetically doped semiconductors and oxides, assigned now to the presence of TM aggregates \cite{Sato:2010_RMP,Bonanni:2010_CSR}.

In this paper we show that this progress in {\em ab initio} and nanocharacterization methods makes it possible to establish the origin of bulk crystalline in-plane uniaxial anisotropy  found in magnetotransport \cite{Katsumoto:1998_PSSB,Tang:2003_PRL,Gould:2008_NJP,Glunk:2009_PRB}, magnetooptical \cite{Hrabovsky:2002_APL,Welp:2003_PRL}, magnetic \cite{Welp:2003_PRL,Welp:2004_APL,Sawicki:2005_PRB,Kim:2007_JKPS}, and ferromagnetic resonance \cite{Liu:2003_PRB} studies of (Ga,Mn)As. We show quantitatively that this puzzling anisotropy, whose presence contradicts the results of group theory for zinc-blende crystals, results from a non-random distribution of Mn over cation sites, setting in at the growth surface during the epitaxy. This insight allows us to propose methods for its controlling, the important step for further exploration of functionalities associated with its presence in (Ga,Mn)As and related systems \cite{Wunderlich:2006_PRL,Chiba:2008_N,Chernyshov:2009_NP,Endo:2010_APL}. Furthermore, our model elucidates the origin of a threefold enhancement of the shape magnetic anisotropy found in thin films of (Ga,Mn)As \cite{Glunk:2009_PRB}.

We consider here zinc-blende (Ga,Mn)As grown by low-temperature molecular beam epitaxy \cite{Ohno:1998_S} along the [001] direction. Under these conditions, long-range aggregation of substitutional Mn cations is kinematically limited, as according to three-dimensional atom probe measurements the Mn distribution is random down to at least 1\,nm \cite{Kodzuka:2009_UM}. Actually, the formation of Mn-rich (Mn,Ga)As nanocrystals inside (Ga,Mn)As films starts to be visible under annealing at temperatures considerably greater than the growth temperature \cite{De_Boeck:1996_APL,Sadowski:2011_PRB}. Thus, we start our studies by finding out the energetically favorable position of the nearest-neighbor (NN) Mn cation dimer on the GaAs (001) {\em surface}, where constituent atoms are mobile during the epitaxy. We perform {\em ab initio} calculations employing  SIESTA code \cite{Ordejon:1996_PRB,Soler:2002_JP}, whose localized basis is well suited for surface studies. The computations within the local-density approximation are carried out for the following geometry: a pair of Mn impurities is located on the $(001)$ surface of a twelve monolayers thick GaAs supercell with lateral dimensions of two lattice constants. The lower Ga-terminated surface is saturated with pseudoatoms of $Z = 1.25$, yielding a total of 112 atoms in a supercell, with a $1.5$~nm thick vacuum region. A $5\times5\times1$  Monkhorst-Pack grid of $k$-points is used, with the slab dipole correction and simulate doping options enabled. Since we are interested in the situation under growth conditions, we have performed a non-spin-polarized calculation with the lattice optimization.

According to the computation results, the preferred orientation of Mn dimers is $[\bar{1}10]$, with the energy gain comparing to the NN $[110]$ pair being as large as $1.0 \, \mathrm{eV}$. Among many consequences of such a non-random Mn distribution is the appearance of a local strain, as Mn atoms in the pair are displaced of the GaAs cation positions, whereas the As atom between them is shifted along the $[00\bar{1}]$ direction by as much as $2.6\%$ of the bond length. The resulting strain may contribute to the formation of stacking faults propagating in the $(111)$ and  $(11\bar{1})$ planes \cite{Ohno:2003_APL}, observed in (Ga,Mn)As by high resolution electron transmission microscopy \cite{Kong:2005_JAP} and synchrotron x-ray diffraction \cite{Kopecky:2011_PRB}. Importantly, according to recent {\em ab initio} studies \cite{Kopecky:2011_PRB}, the intersection lines of the stacking fault pairs may enhance further the aggregation of Mn dimers along the $[\bar{1}10]$ crystallographic directions.

We evaluate the effect of this specific distribution of Mn ions upon magnetic anisotropy in two steps. First, by employing the group theory and the Luttinger's method of invariants we establish the expected form of magnetic energy depending on the magnetization orientation. Second, by making use of either perturbation theory or an {\em ab initio} method we evaluate the magnitude of anisotropy constants brought about by the non-random Mn distribution.

More specifically, we place a nearest neighbor (NN) Mn pair in a GaAs supercell. In order to consider various spatial correlations between these pairs, the supercells are arranged in two ways, corresponding to positions of the Mn dimers in a simple cubic or face-centered cubic sublattice, respectively. Such a system possesses the $C_{2v}$ symmetry. If the Mn pair is along $[110]$ direction, this group  comprises a twofold rotation axis $[001]$ and a reflection plane $(110)$.
Accordingly, the magnetic anisotropy energy $F(m_x, m_y, m_z)$ as a function of the magnetization direction $(m_x, m_y, m_z) = (\sin \vartheta \cos \varphi,
\sin \vartheta \sin \varphi, \cos \vartheta)$ can be expanded into invariants $f_i$,
\begin{equation}
  F = \sum_{i} K_i f_i,
  \label{eq: 1}
\end{equation}
where the basis invariants are defined respecting the decomposition of the space of spherical harmonics with given $l$ into irreducible representations of the $T_{\text{d}}$ group. For example, the decomposition $E \oplus T_2$ of the space of spherical harmonics with $l = 2$ defines invariants $f_1 = m_z^2 - \frac13$ (representation $E$) and $f_2 = m_x m_y$ (representation $T_2$). These are out-of-plane and in-plane uniaxial anisotropies. For $l = 4$, the decomposition
is $A_1 \oplus E \oplus T_1 \oplus T_2$, where the invariant from $A_1$ is the cubic anisotropy, the ones from $E$ and $T_2$ are higher order uniaxial anisotropies analogous to those
with $l = 2$, and there is no invariant in $T_1$.

Let us consider now possible NN cation configurations. Up to the translational symmetry, there are six different placements corresponding to the three planes [\emph{i.e.}, (100), (010), and (001)] in which a pair can be located and two possible directions in each of the planes.  As the growth occurs along the [001] direction, we cannot expect the same density of dimers in the (001) plane comparing to the (100) and (010) planes. Thus, averaging of $f_1$ over the three planes may lead to $K_1 \ne 0$, adding to the effect of epitaxial biaxial strain. Furthermore, by symmetry, the densities of NN pairs in the two directions in the $(100)$ and $(010)$ planes are equal, therefore $f_2$ averages to zero for each of those two planes. However, the symmetry allows a different density of NN pairs along the $[110]$ and $[\bar{1}10]$ directions in the $(001)$ growth surface. This results in a non-zero value of the macroscopic parameter $K_2$. The macroscopic value of~$K_2$ is a product of the anisotropy energy of a single Mn pair and the difference in the densities (per supercell volume) of pairs along the $[110]$ and $[\bar{1}10]$ directions, respectively.

More formally, the correlations in occupations between the neighboring sites form a representation of the point group $T_{\text{d}}$, which decomposes into a sum of the irreducible components, $A_1 \oplus E \oplus T_2$. The irreducible components are the total pair density ($A_1$), the distribution of pairs among the three planes ($E$), and the differences between occupations of the two directions in each of the planes ($T_2$). Similarly, the magnetic anisotropy free energy function decomposes into $E \oplus T_2$ (for $l = 2$) and
$A_1 \oplus E \oplus T_1 \oplus T_2$ (for $l = 4$). Here, we consider a first order (linear) dependence of the magnetic anisotropy on the pair correlations. Hence, the irreducible components are in a direct correspondence: the cubic anisotropy parameter depends linearly on the total pair density (representation $A_1$), the out-of-plane uniaxial anisotropy parameter $K_1$ on the distribution of pairs among the planes ($E$), and the in-plane uniaxial anisotropy constant $K_2$ on the difference between the densities of the $[110]$ and $[\bar{1}10]$ pairs ($T_2$). The remaining component,
$T_1$, has no counterpart in the decomposition of the correlation function and therefore has to vanish. We conclude that the non-random distribution of Mn cations of the form introduced here leads to the functional of the magnetic anisotropy energy (Eq.~\ref{eq: 1}) consistent with experimental results \cite{Tang:2003_PRL,Glunk:2009_PRB,Hrabovsky:2002_APL,Welp:2003_PRL, Welp:2003_PRL,Welp:2004_APL,Sawicki:2005_PRB,Liu:2003_PRB}.

Previously, the anisotropy constant $K_2$ accounting for experimentally observed bulk uniaxial in-plane crystallographic anisotropy was described theoretically by assuming the presence of shear strain, whose magnitude $\varepsilon_{xy}$ was treated as an adjustable parameter \cite{Sawicki:2005_PRB,Zemen:2009_PRB,Stefanowicz:2010_PRBa}, evaluated with the $p-d$ Zener model to be of the order of $\varepsilon_{xy}\approx 0.05$\% for the  relevant GaAs deformation potential $d = -4.8$~eV \cite{Zemen:2009_PRB,Stefanowicz:2010_PRBa}. Such a shear deformation of (Ga,Mn)As films has not been found experimentally \cite{Kopecky:2011_PRB}. Similarly, an additional contribution to $K_1$, found in studies of magnetic anisotropy as a function of biaxial strain \cite{Glunk:2009_PRB}, can be parametrized by $\varepsilon_{xx}\approx -0.05$\%. We will show now that new terms in the Luttinger hamiltonian brought about by the non-random Mn distribution have the form of a strain hamiltonian whose elements have signs and magnitudes consistent with the experimental values.

We first examine how lowering of symmetry to $C_{2v}$ affects the three-band effective mass hamiltonian describing the valence band in the cubic case. The spin-orbit and $p-d$ interactions can then be taken into account in the standard way leading to the six-band hamiltonian from which the magnetic anisotropy energy $F$ (Eq.~\ref{eq: 1}) can be directly determined \cite{Zemen:2009_PRB,Stefanowicz:2010_PRBa}. By employing the Luttinger's method of invariants \cite{Bir:1974_B}, we find that the the presence of spatial correlations of Mn ions is captured by the use of the virtual crystal $k \cdot p$ hamiltonian with terms corresponding to effective shear and biaxial strains, described by two components of the strain tensor,
$\varepsilon_{xy}$ and $\varepsilon_{xx}$, respectively, together with the corresponding deformation potentials $d = -4.8$~eV and $b = -2.0$~eV. The effective biaxial strain $\varepsilon_{xx}$ (with $\varepsilon_{yy} = \varepsilon_{xx}$, $\varepsilon_{zz} = -2 c_{12} \varepsilon_{xx}/c_{11}$, $c_{12}/c_{11} = 0.453$) will renormalize the magnitude of strain coming from a mismatch to the substrate.

\begin{figure}%
\centerline{\hspace{3pt}\rlap{\raisebox{165pt}{a)}}\includegraphics[width=0.945\columnwidth]{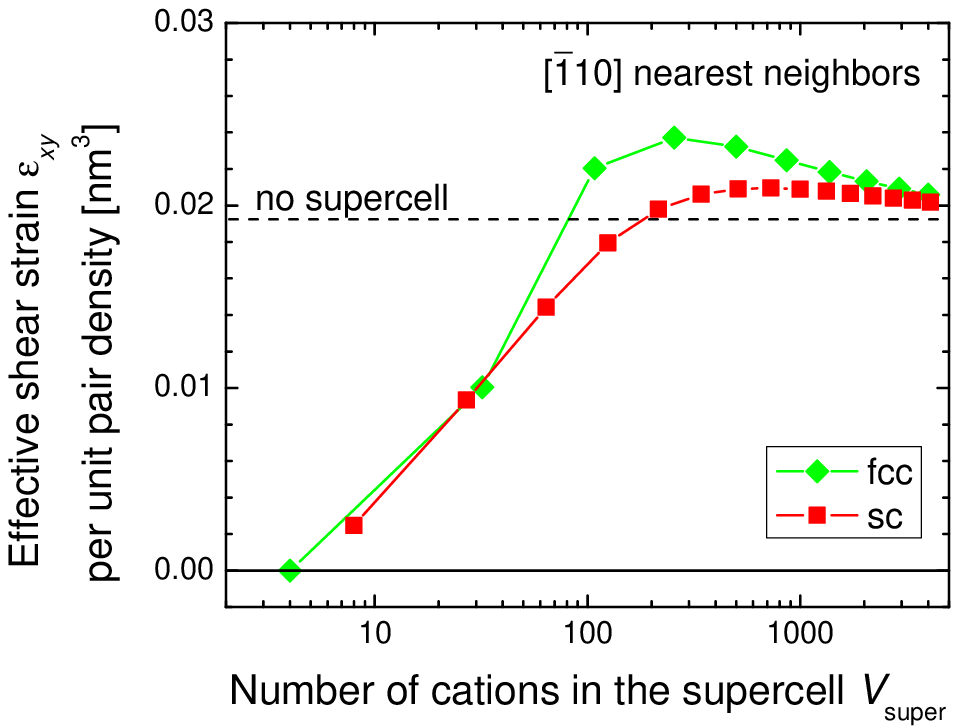}}%
\vspace{12pt}
\centerline{\rlap{\raisebox{165pt}{b)}}\includegraphics[width=0.96\columnwidth]{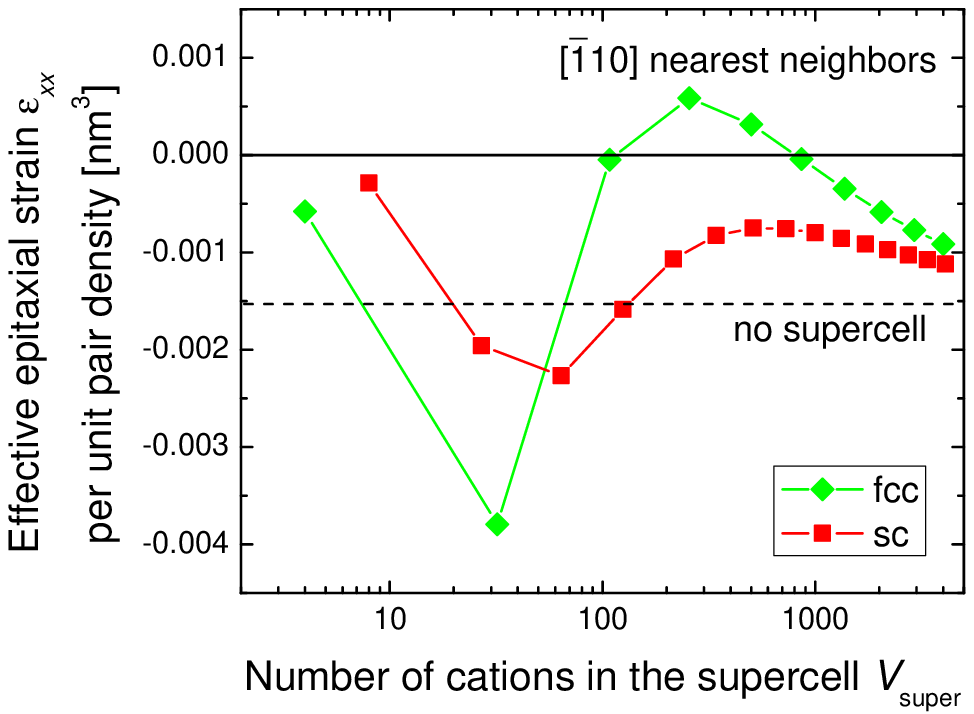}}%
\caption{(Color online) Effective shear $\varepsilon_{xy}$ (a) and epitaxial biaxial $\varepsilon_{xx}$ (b) strains, normalized to unit pair density, for a nearest-cation neighbor $[\bar{1}10]$ pair of Mn ions in supercells arranged in simple cubic and face-centered lattices, as calculated by using Eq.~(\ref{eq: kp pert}).}
\label{fig: 1}%
\end{figure}

In order to evaluate the magnitude of  $\varepsilon_{xy}$ and $\varepsilon_{xx}$, we start from the three-band unperturbed hamiltonian,
\begin{eqnarray}
  \lefteqn{H_{3\times 3}(\mathbf{k}) = E_v \mathbb{1}_{3\times 3} - \frac{\hbar^2}{2m} \times {}} \\
   & & \quad {} \times \left[ \begin{smallmatrix}
    A k_x^2 + B (k_y^2+k_z^2) & C k_x k_y& C k_x k_z\\
    C k_x k_y& A k_y^2 + B (k_x^2+k_z^2)& C k_y k_z\\
    C k_x k_z& C k_y k_z& A k_z^2 + B (k_x^2+k_y^2)
  \end{smallmatrix} \right], \nonumber
\end{eqnarray}
where in terms of the Luttinger parameters $\gamma_i$, $A = \gamma_1 + 4 \gamma_2$, $B = \gamma_1 - 2 \gamma_2$, and $C = 6 \gamma_3$. The Mn potential is assumed to contain a contribution from a screened Coulomb part and a central cell correction, leading in the $k$-space to
\begin{equation}
  \tilde V(k) = - \frac{e^2}{\epsilon \epsilon_0} \frac{1}{k^2 + r_s^{-2}} - \pi^{\frac32} V_0 r_0^3 e^{-k^2 r_0^2/4},
\end{equation}
where the static dielectric constant $\epsilon = 12.9$,
the screening radius is $r_s = 0.5 \, \mathrm{nm}$, and the Gaussian central cell correction parameters are
$r_0 = 0.28 \, \mathrm{nm}$, $V_0 = 3.0 \, \mathrm{eV}$ \cite{Bhattacharjee:2000_SSC}.

Within the second order perturbation theory for a Mn pair we obtain at $\mathbf{k} = 0$,
\begin{eqnarray}
  \Delta H^{(2)}_{3\times 3} & = & g_{\mathbf{d}} \int \frac{d^3 \mathbf{k}}{(2\pi)^3}
    \left(2 \cos \frac{\mathbf{k} \cdot \mathbf{d}}{2} \right)^2 |\tilde V(k)|^2 \times {} \nonumber \\
    & &  \qquad {} \times \left[ E_v \mathbb{1}_{3\times 3} - H_{3\times 3}(\mathbf{k}) \right]^{-1},
    \label{eq: kp pert}
\end{eqnarray}
where $\mathbf{d}$ is the vector between the two Mn ions forming the pair and $g_{\mathbf{d}}$ is the density of such pairs. In the case of a supercell, the integral should be replaced with a sum over the reciprocal lattice $\int \frac{d^3 \mathbf{k}}{(2\pi)^3} \to \frac{1}{V_{sc}} \sum_{\mathbf{k} \ne 0}$, where $\mathbf{k} = 0$ is omitted from the sum and $V_{\text{sc}}$ is the volume of the supercell. The above formula assumes additivity of the Mn potential and neglects distortion of the lattice in the presence of the Mn pair. A similar formula without the squared cosine factor can be used to determine the anisotropy of a single Mn acceptor in a non-cubic supercell \cite{Strandberg:2009_PRB}, as a single Mn acceptor residing in a supercell exhibits a significant magnetic anisotropy in accord with the symmetry of the supercell.

The obtained results are presented in Fig.~\ref{fig: 1} for the NN Mn pair residing along the preferred direction $[\bar{1}10]$. In particular, in the case of no supercell, that is when positions of the $[\bar{1}10]$ dimers can be regarded as uncorrelated, we obtain  $\varepsilon_{xy}/g_{\mathbf{d}} = 0.019 \, \mathrm{nm}^3$, $\varepsilon_{xx}/g_{\mathbf{d}} = -0.0015 \, \mathrm{nm}^3$.  For the Mn concentration $x =6.25\%$, this leads to $\varepsilon_{xy} = 1.33\%$, $\varepsilon_{xx} = -0.10\%$. However, when the positions of the $[\bar{1}10]$ dimers are correlated, $\varepsilon_{xy} = 0.77\%$, $\varepsilon_{xx} = -0.16\%$ in the case of a 54 atoms fcc supercell and $\varepsilon_{xy} = 0.70\%$, $\varepsilon_{xx} = -0.26\%$ in the case of a 64 atoms sc supercell, corresponding to (Ga,Mn)As with 7.41\% and 6.25\% of Mn arranged into a regular sublattice of $[\bar{1}10]$ Mn dimers.

Comparing to experimental findings, we see that the computed values of $\varepsilon_{xy}$ and $\varepsilon_{xx}$ have the correct sign. At the same time, their absolute values are significantly greater that the one determined experimentally, $\varepsilon_{xy}\approx 0.05$\% and $\varepsilon_{xx}\approx -0.05$\%. This could be expected as in real samples only a fraction of the Mn content forms NN pairs as well as some Mn pairs choose the less preferred direction during the growth process.

It is instructive to compare the above findings to {\em ab initio} result. For this calculations particularly suitable is the Quantum ESPRESSO code developed in the plane wave basis \cite{Giannozzi:2009_JPCM}. Within local spin density approximation and for supercells with 54 cations (fcc) and 64 cations (sc), using a $4\times4\times4$ grid of $k$-points, we obtain from magnitudes of the valence band splitting at the $\Gamma$ point of the Brillouin zone,
$\varepsilon_{xy} = 1.70\%$ ($1.79\%$), $\varepsilon_{xx} = -0.75\%$ ($-1.01\%$) and $\varepsilon_{xy} = 1.37\%$ ($1.50\%$), $\varepsilon_{xx} = -0.49\%$ ($-0.47\%$) without (with) lattice optimization, respectively.

In summary, we argue that puzzling bulk in-plane crystalline magnetic anisotropy in (001) (Ga,Mn)As is brought about by the preferred formation of Mn dimers along the $[\bar{1}10]$ at the growth surface, as implied by the {\em ab initio} results. The group theory and the Luttinger's method of invariants as well as the perturbation and {\em ab initio} computation results show that the effect of the predicted Mn distribution can be parametrized by effective shear and biaxial strains. Their signs agree with the experimental determination while the computed absolute values are much larger than the experimental magnitudes indicating that the surplus of Mn dimers residing in the preferred positions is only partial after the growth process.  This suggests that it might be possible to control the strength of uniaxial anisotropy by changing the epitaxy conditions, particularly the growth rate and/or temperature. Furthermore, when reducing the film thickness, interfacial and surface anisotropies, for which in-plane uniaxial anisotropy is allowed even for a random distribution of magnetic ions \cite{Sawicki:2004_PRB,Mankovsky:2011_PRB}, may gradually come into play. In general terms, our results show how a specific microscopic distribution of alloy constituents may affect the symmetry properties and magnitudes of macroscopic response functions.

The work was supported by FunDMS Advanced Grant of ERC (Grant No. 227690) within the Ideas 7th Framework Programme of European Community, InTechFun (Grant No.
POIG.01.03.01-00-159/08), and SemiSpinNet (Grant No. PITNGA-
2008-215368). We have used the computing facilities of PL-Grid Polish Infrastructure for Supporting
Computational Science in the European Research Space and ICM Interdisciplinary Centre for Mathematical and Computational Modelling, University of Warsaw.

\bibliography{uni110}

\end{document}